\documentclass{article}
\pdfoutput=1
\usepackage{smc2020}
\usepackage{times}
\usepackage{ifpdf}
\usepackage[english]{babel}
\usepackage{cite}
\usepackage{booktabs}
\usepackage[redeflists]{IEEEtrantools}
\usepackage{arydshln}

\def\papertitle{TOWARDS RELIABLE REAL-TIME OPERA TRACKING:\break COMBINING ALIGNMENT WITH AUDIO EVENT DETECTORS TO INCREASE ROBUSTNESS}
\def\firstauthor{Charles Brazier}
\def\secondauthor{Gerhard Widmer}

\newif\ifpdf
\ifx\pdfoutput\relax
\else
   \ifcase\pdfoutput
      \pdffalse
   \else
      \pdftrue
\fi

\ifpdf 
  \usepackage[pdftex,
    pdftitle={\papertitle},
    pdfauthor={\firstauthor, \secondauthor},
    bookmarksnumbered, 
    pdfstartview=XYZ 
   ]{hyperref}

  \usepackage[pdftex]{graphicx}
  \graphicspath{{./figures/}}
  \DeclareGraphicsExtensions{.pdf,.jpeg,.png}

  \usepackage[figure,table]{hypcap}

\else 
  \usepackage[dvips,
    bookmarksnumbered, 
    pdfstartview=XYZ 
  ]{hyperref}  

  \usepackage[dvips]{epsfig,graphicx}
  \graphicspath{{./figures/}}
  \DeclareGraphicsExtensions{.eps}

  \usepackage[figure,table]{hypcap}
\fi

\hypersetup{
    colorlinks,%
    citecolor=black,%
    filecolor=black,%
    linkcolor=black,%
    urlcolor=black
}

\title{\papertitle}

\twoauthors
   {\firstauthor} {Institute of Computational Perception \\ Johannes Kepler University Linz, Austria\\ %
     {\tt \href{mailto:charles.brazier@jku.at}{charles.brazier@jku.at}}}
   {\secondauthor} {Institute of Computational Perception and \\
                    LIT AI Lab, Linz Institute of Technology \\
                    Johannes Kepler University Linz, Austria \\ %
     {\tt \href{mailto:gerhard.widmer@jku.at}{gerhard.widmer@jku.at}}}

\begin{document}
\bstctlcite{IEEEexample:BSTcontrol}
\capstartfalse
\maketitle
\capstarttrue

\begin{abstract}
Recent advances in real-time music score following have made it possible for machines to automatically track highly complex polyphonic music, including full orchestra performances. In this paper, we attempt to take this to an even higher level, namely, live tracking of full operas. We first apply a state-of-the-art audio alignment method based on online Dynamic Time-Warping (OLTW) to full-length recordings of a Mozart opera and, analyzing the tracker's most severe errors, identify three common sources of problems specific to the opera scenario. To address these, we propose a combination of a DTW-based music tracker with specialized audio event detectors (for applause, silence/noise, and speech) that condition the DTW algorithm in a top-down fashion, and show, step by step, how these detectors add robustness to the score follower. However, there remain a number of open problems which we identify as targets for ongoing and future research.
\end{abstract}

\section{Introduction}
\label{sec:introduction}

Score following is the task of aligning a music performance with its corresponding score, in real time, in order to be able to follow along in the score and synchronize the live music with various other media or applications (such as score viewers, page turners \cite{arzt2008automatic}, lighting, etc.). Since its start in the 1980s \cite{dannenberg1984line}, score following research has produced promising results, eventually succeeding in following full orchestral performances in the concert hall  \cite{prockup2013orchestral, arzt2015real}, and it is still an active research topic in Music Information Retrieval \cite{munoz2019new, vandermethod}. The goal of our project is to extend this to even more complex musical stage works, namely, entire \textit{live operas}. Robust real-time opera tracking would have many applications in the opera hall as well as in live streaming scenarios, such as fully automatic lyrics displays, synchronization of lighting, camera control, live video cutting, and back-stage operations. In fact, our partner in this project is the \textit{Vienna State Opera}, a world-famous opera house and one of the pioneers in HD and UHD opera live streaming\footnote{https://www.wiener-staatsoper.at/en/}.

Two state-of-the-art methods are commonly mentioned in the score following literature. One is a probabilistic approach where the process is modeled by a Hidden Markov Model \cite{raphael1999automatic, raphael2006aligning, cont2009coupled}, the current position being the hidden variable and the audio performance being the observation. The other approach makes use of \textit{Dynamic Time Warping (DTW)} \cite{arzt2008automatic, macrae2010accurate} to find the optimal alignment between score sequence and performance sequence.

\begin{table*}[t]
 \begin{center}
 \begin{tabular}{@{}llllcl@{}}
  \toprule
  \textbf{Conductor} & \textbf{Orchestra} & \textbf{Place} & \textbf{Year} & \textbf{Duration} & \textbf{Role}\\
  \midrule
  \'A. Fischer & Orchester der Wiener Staatsoper & Vienna State Opera, Vienna, AT & 2018 & 3:12:54 & Target \\
  Karajan & Berliner Philharmoniker & Philharmonie, Berlin, DE & 1985 & 2:57:53 & Reference \\
  \bottomrule
 \end{tabular}
\end{center}
 \caption{Dataset used in the audio-to-audio alignment strategy.}
 \label{tab:dataset}
\end{table*}

Synchronizing two different entities requires a common representation. In musical score following, the score is generally converted into a corresponding symbolic representation (MIDI or MusicXML format) and then synthesized into audio, using appropriate sound fonts. The task is then to align two audio streams (or feature sequences extracted from these) to each other, in real time. The problem with this is that the resulting score audio can be very different from the performance in terms of sound, but also in terms of structure. This becomes particularly acute in the domain of opera, with its complex structure (including unexpected events), and the complexity of voices and singing techniques of opera singers. (Also, it is next to impossible to obtain full opera scores in machine-readable format.)

An alternative is to use some available real \textit{audio recording(s)} of the piece in question, annotated and aligned to the score, as a proxy of the score, and treat score following as a performance-to-performance matching problem. This approach has recently been demonstrated to successfully follow a complete orchestra concert \cite{arzt2015real}. In this paper, we will follow this latter strategy, presenting a tracker that attempts to align a live opera performance (the \textit{target}) with another, already recorded and annotated performance (the \textit{reference}), using an adapted version of the \textit{On-Line Dynamic Time Warping (OLTW)} algorithm \cite{dixon2005line}.

However, as we will see, operas are generally more complex than orchestral concerts, containing a complex mixture of voices (singing voice or spoken language) and music which is often interrupted by breaks, noises, interludes, etc. A tracking experiment with a full annotated Mozart opera (\textit{Don Giovanni}) will reveal several general sources of problems. For three of these -- applause, acting breaks, and non-notated musical interludes --, we will propose a solution based on learning specific acoustic event detectors and using these to control the DTW tracking process. A series of experiments will demonstrate the additive beneficial effect of these detectors.

The paper is structured as follows. Section \ref{sec:data_description} describes the data provided by the Vienna State Opera, as well as the manual annotations needed for evaluation. Section \ref{sec:standard_OTW} demonstrates the limitations of a standard online alignment system. From an analysis of the most severe tracking errors, Section \ref{sec:challenges} derives some classes of problematic situations. Section \ref{sec:model} describes our acoustic event detectors, how they were trained, and how they interact with the OLTW tracker. Section \ref{sec:experiments} presents corresponding experiments, Section \ref{sec:discussion} discusses the results, and Section \ref{sec:conclusion} offers some conclusions and ideas for the next research steps.

\section{Data description}
\label{sec:data_description}

For systematic experiments with opera score following, annotated audio data are needed. There is as yet no such data set, so we had to produce our own. As a first test piece, we have selected the opera ``Don Giovanni" by W.A.Mozart.\footnote{We are aware of the fact that a single opera (in two different recordings) may seem like rather little data. However, it should be kept in mind that a full opera represents several hours of music and requires massive annotation work. For instance, the score of `Don Giovanni' is a book of 500 pages, containing 5304 bars of music. Efforts are currently underway to annotate additional Don Giovanni recordings, and a second Mozart opera (Die Zauberfl\"ote; again in several recordings), in order to enlarge the current study.}  \tabref{tab:dataset} summarises the performances/recordings used. For our project, we were provided with the recording of a live performance at the Vienna State Opera in Nov.~2018, conducted by \'Ad\'am Fischer, in the form of one continuous audio recording of almost 4~hours. This is the \textit{target} that we want to track relative to the score. To slightly simplify the task, we do not consider the extended audio passages before the beginning of the acts, where musicians rehearse or tune their instruments while the audience arrives in the opera hall.

As a \textit{reference} and proxy for the score, we chose a CD recording by Herbert von Karajan from 1985 (Deutsche Grammophon). This performance is split into tracks on the CD. We noted that a few specific tracks, e.g., Act 1, Scene 6, Recitativo (\textit{In questa forma dunque}) and Act 2, Scene 12, Recitativo (\textit{Ah, si segua il suo passo}), are present in the reference but were omitted in the target performance. As a first simple solution, we decided to remove these from the reference (Karajan) in order to obtain the same musical structure of reference and target. Later, in Section \ref{sec:discussion}, we will briefly sketch how the problem of structural mismatch could be handled.

We manually annotated both recordings by placing time markers at bar boundaries, giving us 5304 bar annotations (2866 for the first act, 2438 for the second) for each recording. The annotations made on the reference performance permit us to link the audio to the score, while the annotations made on the target serve as ground truth and allow us to evaluate the accuracy of the tracker.

For evaluation, the metric described in \cite{cont2007evaluation} is applied. For each alignment given by the tracker, we compute its errors with respect to the ground truth and calculate the mean and standard deviation of the alignment errors, as well as the percentage of bar boundaries matched with an error below one, two, and five seconds. We also provide the maximal alignment error $err_{max}$ (in seconds).

\begin{figure*} [t]
\centering
\includegraphics[width=0.8\textwidth]{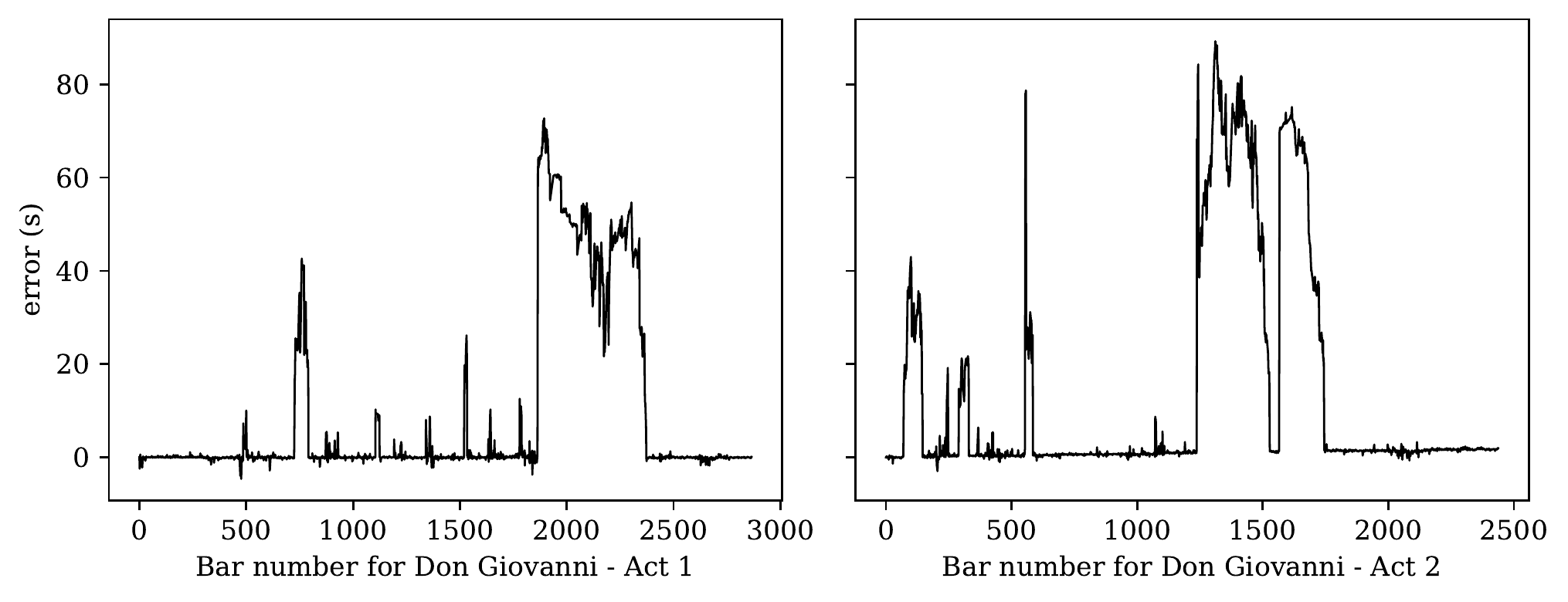}
\caption{Bar-level alignment error on Don Giovanni, Act I (left) and Act II (right).\label{fig:OTW_errorcurve}}
\end{figure*}

\begin{table} [t]
 \begin{center}
 \begin{tabular}{@{}lrrrrrr@{}}
  \toprule
  \textbf{} & \textbf{Mean} & \textbf{Std} & $\mathbf{\leq 1s}$ & $\mathbf{\leq 2s}$ & $\mathbf{\leq 5s}$\ & \textbf{$err_{max}$}\\
  \midrule
  \textbf{Act 1} & 9.6s & 19.1s & 0.74 & 0.76 & 0.78 & 72.7s \\
  \textbf{Act 2} & 14.0s & 24.7s & 0.40 & 0.72 & 0.74 & 89.2s \\
  \bottomrule
 \end{tabular}
\end{center}
 \caption{Standard OLTW: Tracking error on complete acts. (Absolute durations in target: Act 1: 01:38:13; Act 2: 01:34:41)}
 \label{tab:OTW_table}
\end{table}

\section{Starting Point: OLTW-based Tracking}
\label{sec:standard_OTW}

Standard Dynamic Time-Warping (DTW) \cite{muller2007dynamic} calculates the optimal alignment path between two sequences. However, the inherently quadratic complexity of the similarity/cost matrix computation is prohibitive for online or real-time applications. In our work, we thus opted for the \textit{On-Line Time Warping (OLTW)} algorithm described in \cite{dixon2005line}, which is of linear complexity, and adapt it to our purpose. For real-time application, each element of the \textit{target} sequence received incrementally is compared to a vector of elements in the \textit{reference}, which is centered around the expected position in the reference sequence. This vector can be interpreted as a calculation security margin around the position proposed by the algorithm. The size of the vector directly conditions the calculation time. For our experiments, the length of the vector has been fixed to correspond to 40~seconds of audio.

In order to align two audio recordings, we first need to select suitable audio features to be compared. Following the study in \cite{gadermaier2019study}, we extract, from each audio frame, 120~MFCCs but discard the first 20, which leaves us with a feature vector of length 100. The audio length is fixed to 20~ms with a hop size of 10~ms. With this set of parameters, we compute the features of the \textit{reference} performance beforehand, whereas the features from the \textit{target} are computed in real time.

In accordance with the same study, we select the cosine distance to measure the cost (difference) between the current target feature vector and a range of feature vectors (the above-mentioned `calculation security margin') in the reference. From the resulting cost vector, we recursively compute the corresponding global cost in following the standard accumulated cost formula (see equation (4.5) in \cite{muller2007dynamic}). We finally retrieve the score position by extracting the index minimizing the global cost.

Fig.~\ref{fig:OTW_errorcurve} shows the error curves (time deviations between alignment and ground truth) when we perform OLTW on each of the two acts making up Don Giovanni. A positive error means that the tracker is in advance in the score, while a negative error indicates a delay. The corresponding numeric results given in Table \ref{tab:OTW_table} constitute our baseline. Of the 2866~annotations (bars) making up Act 1, 74\% have been detected with an error smaller or equal to 1~second, with a maximum error of 72.7~s. Analogously, for Act 2, composed of 2438~annotations, only 40\% have been detected with an error smaller or equal to 1~second, with a maximum error of 89.2~s. These results as well as the curves in Fig.~\ref{fig:OTW_errorcurve} show that the tracker has severe problems. A positive aspect is that it does not get completely lost; it does eventually manage to find back to the correct position in the reference, but sometimes only after a long time (note, e.g., the period of disorientation in Act 1, which lasts around 500 bars!). Our goal in the following is to identify the main sources of difficulty and reasons for some of these errors, and propose ways to address these.

\section{Some Opera-specific Challenges}
\label{sec:challenges}

As a first step, we take a look at those points in the tracking error curves (Fig.~\ref{fig:OTW_errorcurve}) where the error suddenly increases, by a large amount. In particular, there are four anchor points with sudden error jumps higher than 40 seconds\footnote{The two additional clearly visible large error peaks (around bar~730 in Act~1, and near the beginning of Act~2), on analysis, turn out to relate to similar musical/acoustic causes as the four cases we picked for discussion.}. We now take a closer look at these, in order of severity.

The biggest jump occurs during the transition from Act~2, Scene~10, Aria (\textit{Mi trad\`i quell'alma ingrata}) and the following recitative (\textit{Ah ah ah ah, questa \`e buona}) in Scene~11, with the jump equal to 66.5~seconds. In listening to the \textit{target} performance, we find that the end of the aria is followed by 15~s of applause and 54~s of relative silence (no music, no speech) where coughs from the audience, knocks at a door and footsteps are clearly heard.

The second biggest jump (64.3~s) appears during the transition between the \textit{recitativo accompagnato} in Act~2, Scene~12 (\textit{Crudele! - Ah no, mio bene!}) and the start of the Finale. Listening to the live performance, we find the same applause (14~s) -- silence (17~s including coughs and footsteps) sequence as before.

The third biggest jump (60.3~s) happens at the transition from the aria in Act~2, Scene~6 (\textit{Vedai, carino}) to the following recitative (\textit{Di molte faci il lume}). In listening to the target performance we find a more difficult sequence including 12~s of applause, 51~s of silence including coughs, followed by a 17~second improvised interlude played by a cembalo. This interlude is not written in the score and does not appear in the Karajan reference performance.

The fourth big jump of the error curve (46.2~s) appears in Act~1, at the transition between Aria~10a (\textit{Dalla sua pace}, Scene~14, Vienna Version) and the following recitative (\textit{Io deggio ad ogni patto}, Scene~15). We find again the sequence applause (26~s), silence including coughs (24~s), and an interlude (18~s) which is also not written in the score and does not appear in the reference. As the error curve shows, the tracker is then confused for a long period (almost 500~bars) before finding back into track.

From the above, we can immediately identify three specific sources of problems for any opera tracker: (1)~spontaneous applause from the audience; (2)~more or less extended passages of silence or noise (due, e.g., to acting, singers moving about on stage, or stage changes); and (3)~short musical interludes that are not part of the score (often improvised, e.g., on a clavichord or cembalo).\footnote{An additional and general source of problems are \textit{recitativo} sections, where again we see relatively large alignment errors, due to the larger sonic variations and differences between different recordings and the often speech-like character of the singers' parts. This is a problem that requires deeper analysis and more elaborate methods, and is the focus of ongoing work.} In the following, we will propose solutions to these problems by devising specific acoustic event class detectors and integrating these into the tracking procedure.

\section{Combining OLTW with \break Acoustic Event Detectors}
\label{sec:model}

The acoustic event detectors will be based on trained audio classifiers for applause, speech, and music recognition. There are several conceivable ways of combining the detectors with the OLTW algorithm; the simplest one is to use them to directly control the tracking process, halting the tracker during events or passages that are not thought to be part of the score. (Note that all the severe errors in Fig.~\ref{fig:OTW_errorcurve} are positive, meaning that the tracker was ahead of where it should have been.) The model is illustrated in Fig.~\ref{fig:model}. The three specific detectors and corresponding control strategies are as follows:

\begin{figure}[t]
\centering
\includegraphics[width=0.9\columnwidth]{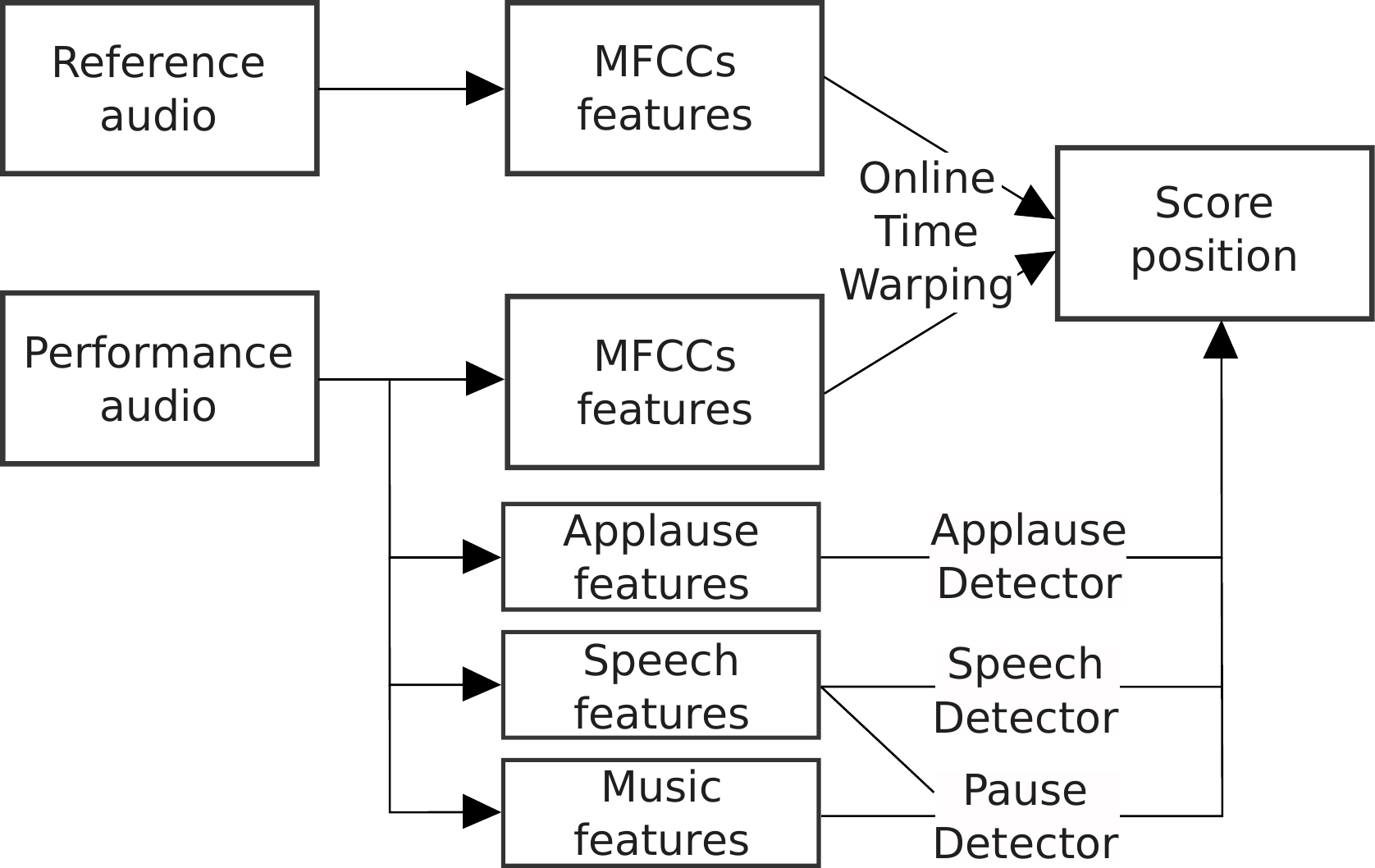}
\caption{Integrated Model.\label{fig:model}}
\end{figure}

\begin{enumerate}

\item \label{applause_strategy}
The purpose of the \textit{Applause Detector} is to control the tracker's advance in the case of spontaneous applause at transitions from one part to the next, i.e., at the end of an aria. The Karajan performance acting as reference does not include applause, but the Fischer live recording does. The error curves in Fig.~\ref{fig:OTW_errorcurve} indicate that when the audience applauds between scenes, the tracker keeps progressing in the score. Our simple strategy is to stop this progression when we detect applause, and to force the current score position to be the transition position (i.e. the beginning of the next part) in the score. In order not to be sensitive to clapping sounds that can be part of some acting action during a scene, this condition is activated only when the current score position is within an interval of 1~s around a transition position according to the score, and when we detect applause for a duration longer than 400~ms, reducing the sensitivity to noise.

\item \label{pause_strategy}
The purpose of the \textit{Pause Detector} is to address the second challenge: extended passages of silence or acting noise between parts. We model these as the periods that contain neither speech nor music (but possibly other noises), using dedicated speech and music classifiers (see below). Our observation is similar to the previous point: the tracker keeps progressing in the score during these passages. Our strategy is to stop this progression when we detect neither speech nor music, again forcing the current score position to be the transition position in the score. As before, this condition is active when the current score position is within an interval of 1~s around a transition position and when we detect pause for longer than 400~ms.



\item \label{interlude_strategy}
A possible approach to the \textit{Interlude Problem} would be to continuously compare, during regular tracking, the input audio stream to the beginning of the next expected section (aria, recitative, etc.) in the reference (via \textit{stream monitoring}~\cite{sakurai2007stream}) so as to be able to re-synchronise at section beginnings. However, this requires to set a threshold on DTW's global cost in order to judge if there is a match or not. In experiments, this turned out to be very brittle. For now, we instead settled on a simple heuristic solution, based on the observation that, at least in operas from the period in question, such improvised interludes usually occur before \textit{recitativo} sections, which themselves are difficult to track (as noted in Section~\ref{sec:challenges}, Footnote~3).
Our heuristic aims at synchronizing the first voice activity of a song between target and reference, if the reference contains voice in the first 4~s of audio (the idea here is to only target \textit{recitativo} sections). When the tracker proposes a score position after a section transition that involves voice activity, the tracker is halted until voice activity is also detected in the live target. This indirect solution also allows the tracker not to accumulate error in the future.
\end{enumerate}

To implement the above control strategies, we learn three dedicated audio classifiers: an applause detector, a music detector, and a speech detector. For training, we use an internal data set consisting of a collection of YouTube recordings of classical music performances. The data set contains 15~hours of audio and was manually annotated with applause, music and speech labels.

In selecting features for training these classifiers, we follow the relevant literature and rely on the features used in the respective current state-of-the-art models, keeping a focus on those that can be computed in real time. For each feature, we fix the window size at 100~ms of audio and the hop size at 20~ms. The features for the applause detector are described in \cite{uhle2011applause} and are a combination of the Spectral Centroid, Spectral Spread, Spectral Flux and Spectral Flatness measures from four different frequency bands, and the first 9 MFCCs from a filter bank of 20 channels. This results in a vector of length 25. The features for the music classifier are a concatenation of the applause features and the \textit{Continuous Frequency Activation (CFA)} feature described in \cite{seyerlehner2007automatic}, giving a vector of length 26. The speech features are a concatenation of the four spectral measures from the applause feature, the \textit{Curved Frequency Trajectory (CFT)} used in speech detection \cite{sonnleitner2012simple}, and the \textit{Fluctogram} from 11 overlapping frequency bands with 18 delta-MFCCs used in singing voice detection \cite{lehner2018online}, resulting in a vector of length 46.

The classifier models are identical and inspired by \cite{lehner2018online}, II.D., and are LSTM-RNNs. Considering their real-time application, they are composed of a unique LSTM layer of 55 cells, corresponding to a temporal context of 1.1~s, followed by a linear layer and a sigmoid function. We train these models with a Binary Cross-Entropy loss function.

\section{Experiments}
\label{sec:experiments}

\begin{figure*}
\centering
\includegraphics[width=0.8\textwidth]{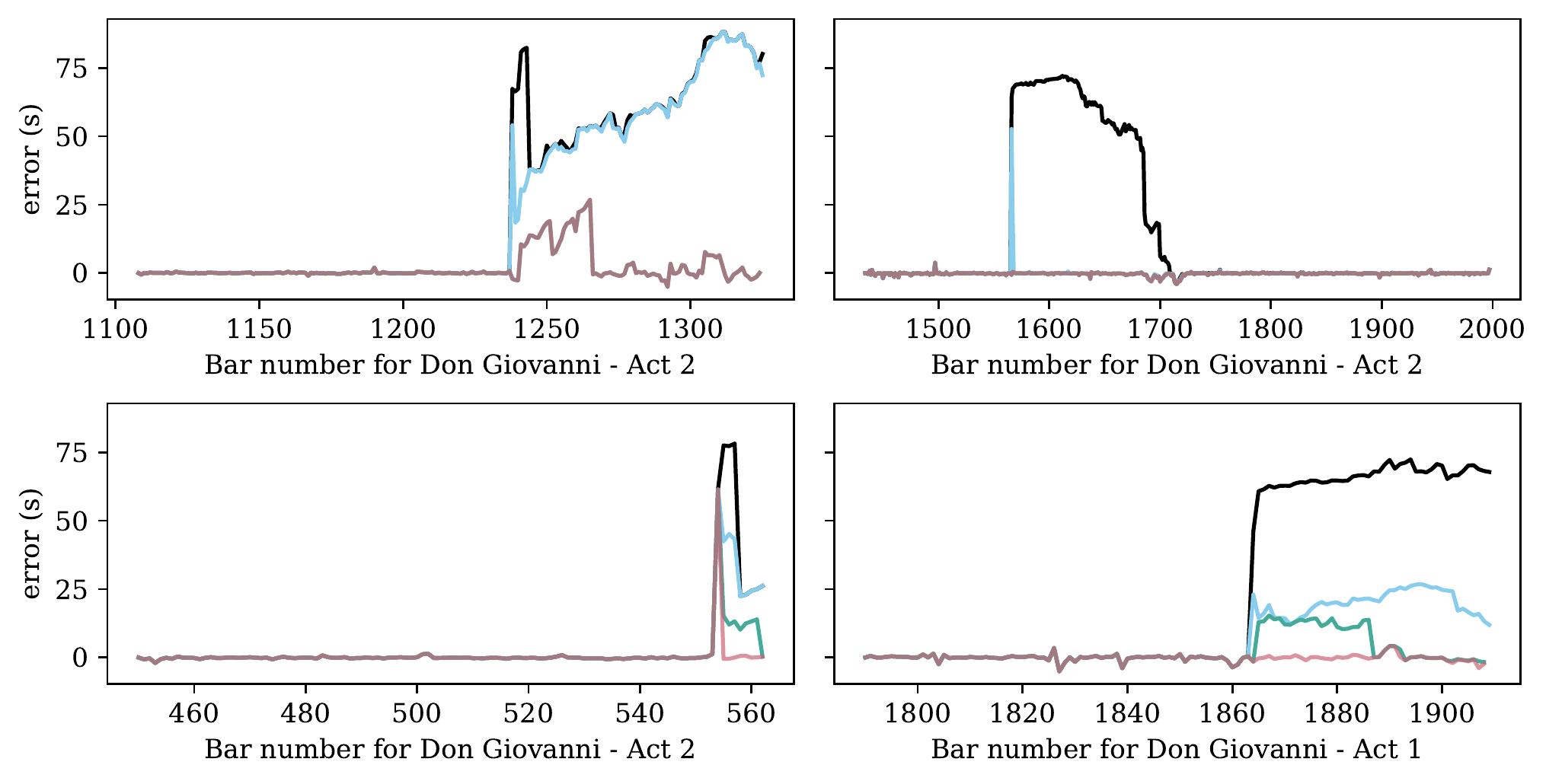}
\caption{Bar-level alignment errors on Don Giovanni, Jump 1 (top-left), Jump 2 (top-right), Jump 3 (bottom-left) and Jump 4 (bottom-right). Black: On-Line Time Warping (OLTW). Blue: OLTW with applause detector. Green: OLTW with applause and pause detectors. Red: OLTW with applause, pause and speech-based interlude detectors. \label{fig:OTW_detectors}}
\end{figure*}

For the following experiments, we add the different detectors incrementally. Starting from the baseline model, we first show the relevance of integrating the Applause Detector, then add the Pause Detector and, finally, the speech-based Interlude Detector. Each experiment is applied specifically on the four problematic sections discussed in detail in Section~\ref{sec:challenges} above. The resulting error curves are illustrated in Fig.~\ref{fig:OTW_detectors}. The full tracking performance on the entire opera acts is summarised in Table~\ref{tab:OTW-detectors_table}.

\begin{table}[t]
 \begin{center}
 \begin{tabular}{@{}lrrrrrr@{}}
  \toprule
  \textbf{} & \textbf{Mean} & \textbf{Std} & $\mathbf{\leq 1s}$ & $\mathbf{\leq 2s}$ & $\mathbf{\leq 5s}$\ & \textbf{$err_{max}$}\\
  \midrule
  \textbf{Act 1} & 9.6s & 19.1s & 0.74 & 0.76 & 0.78 & 72.7s \\
  \textbf{A1+A} & 1.5s & 5.4s & 0.89 & 0.92 & 0.94 & 42.6s \\
  \textbf{A1+AS} & 0.7s & 2.9s & 0.92 & 0.95 & 0.97 & 27.5s \\
  \textbf{A1+ASI} & 0.3s & 0.9s & 0.94 & 0.97 & 0.99 & 12.5s \\ \hdashline
  \textbf{Act 2} & 14.0s & 24.7s & 0.40 & 0.72 & 0.74 & 89.2s \\
  \textbf{A2+A} & 8.7s & 20.7s & 0.79 & 0.82 & 0.83 & 88.2s \\
  \textbf{A2+AS} & 2.4s & 7.5s & 0.87 & 0.90 & 0.92 & 61.5s \\
  \textbf{A2+ASI} & 1.4s & 5.3s & 0.89 & 0.93 & 0.95 & 61.5s \\
  \bottomrule
 \end{tabular}
\end{center}
 \caption{OLTW tracking performance on complete acts (Act~1, Act~2), with applause detector (A1+A, A2+A), with applause and pause detectors (A1+AS, A2+AS) and with applause, pause and speech-based interlude detectors (A1+ASI, A2+ASI).}
 \label{tab:OTW-detectors_table}
\end{table}

\subsection{OLTW with Applause Detector}
\label{sec:OTW_applause}

We first integrate our Applause Detector in the tracking process. This configuration leads to the blue error curves in Figure~\ref{fig:OTW_detectors}. Comparing this to the black curves representing our baseline, we observe an improvement in all of our four cases. We also observe that the size of the initial error jump does not always decrease. This indicates that we do not catch the error immediately during the changeover to the next scene. However, we always observe a drop at the next bar, which means that the tracker quickly recuperates.

The quantitative results on the complete acts are summarised in Table~\ref{tab:OTW-detectors_table}. We see that for both acts, A1+A and A2+A perform much better than Act~1 and Act~2, with 89\% and 79\%, respectively, of the bars aligned with an error below one second.

\subsection{OLTW with Applause and Pause Detectors}
\label{sec:OTW_pause}

In the second experiment, we add the pause detection strategy (based on music and speech detectors), in order to handle possible acting breaks between songs. This configuration corresponds to the green curve in Figure~\ref{fig:OTW_detectors}. We observe that in all four cases, the error decreases, even tracking without any error jump in the second challenge (see top-right).

The overall improvement on the complete acts can be seen in Table \ref{tab:OTW-detectors_table} (columns +AS), with now 92\% and 87\% of the bars from Act~1 and Act~2, respectively, detected below one second.

\subsection{OLTW with Applause \& Pause Detectors and Condition on Speech Activity}
\label{sec:speech_detector}

In the third experiment, we make use of the speech detector to deal with interludes by synchronising the beginnings of voice activity (Strategy~3 above). This results in the red curve in Figure~\ref{fig:OTW_detectors}. We observe further improvement in the latter two of the challenges (see bottom-left and bottom-right in Fig.~\ref{fig:OTW_detectors}), which contain interludes between scenes of 17 and 18~s, respectively. The fourth alignment (see bottom-left) does not present any error jump any more. For the two first cases without interlude, the error curves remain the same.

On the complete acts, Table \ref{tab:OTW-detectors_table} shows that the +ASI variant yields an additional improvement, making it the best alignment system among all of the methods.

\begin{figure*}[t]
\centering
\includegraphics[width=0.8\textwidth]{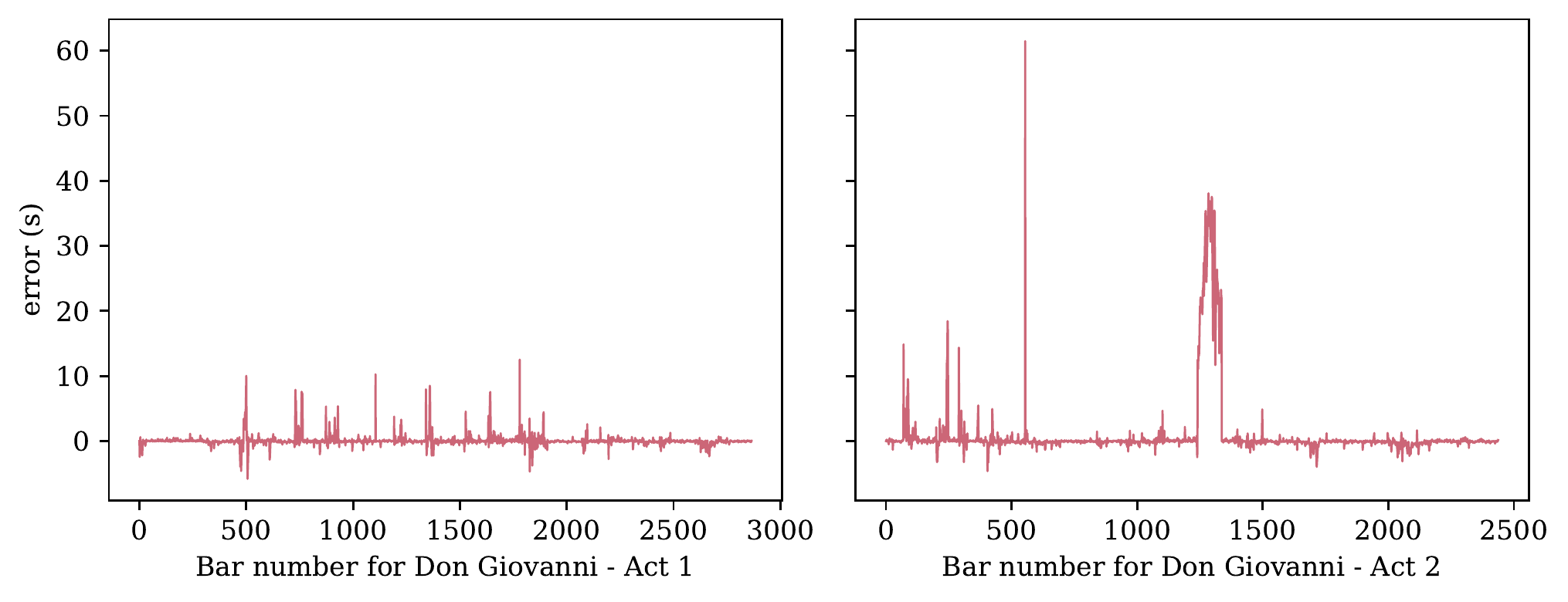}
\caption{Bar-level alignment error on Don Giovanni, Act I (left) and Act II (right). Black: On-Line Time Warping (OLTW). Red: OLTW with applause, pause and speech-based interlude detectors. \label{fig:OTW_error_final}}
\end{figure*}

\section{Discussion of Results and \break Remaining Problems}
\label{sec:discussion}

The complete picture is shown in Fig.~\ref{fig:OTW_error_final}, which gives the error curves of the final full-act alignments produced by our tracker. As can be seen, the presented strategies improved the tracking not only in the four cases we had selected for analysis, but also in some of the other passages that had suffered from severe error (compare this to Fig.~\ref{fig:OTW_errorcurve}). However, there remain a number of open problems.

First, we note that in the second act, the maximal error did not decrease significantly (see also Table~\ref{tab:OTW-detectors_table}). This comes from the detector's condition which aims at forcing the current score position to be the transition position in the score (see \ref{sec:model}.\ref{applause_strategy} and \ref{sec:model}.\ref{pause_strategy}). As a consequence, the time error between this position and its corresponding ground truth can remain large, but this method does help avoid the propagation of the error into the following bars, a problem which had shown as a plateau in the baseline error (see Fig.~\ref{fig:OTW_detectors}).

Looking at the selected situations in Fig.~\ref{fig:OTW_detectors}, we find an error that still persists, visible in the top-left plot. Listening to the audio reveals that there is a second interlude played in between the recitative \textit{Ah ah ah ah, questa \`e buona} (Act~2, Scene~11), making our solution ineffective. This indicates that our current heuristic solution
to the interlude problem is still unacceptably brittle.

The tracking curves for the full acts (Fig.~\ref{fig:OTW_error_final}) reveal that a few problems persist, but almost all of them are corrected very quickly. What is new relative to Fig.~\ref{fig:OTW_errorcurve} are some occasional downward spikes in the error curve. These are due to the activation of our speech-based Interlude Detector which stops the tracking. Looking at the voice activity in our ground truth, which is automatically annotated by our speech detector beforehand, we observe that the signal contains noise which tends to halt the tracker more often than expected. We plan to annotate the reference manually in order to study this problem in more detail.

Generally, \textit{recitatives} are challenging to detect the exact beginning of, and to track. They tend to be played in a different way from performance to performance. Actors pronounce their lyrics in whispering, speaking or singing while they act. Thus the vocal content varies significantly between recordings, making the alignment inaccurate. Also, the accompaniment is partly improvised and even played by different instruments in different recordings. It may thus become necessary to consider the actual lyrics for robust tracking. Consequently, we are currently working towards speech recognition and audio-to-lyrics transcription/alignment methods specifically adapted for opera.

A final problem that needs to be addressed concerns \textit{structural mismatches} between score (reference) and performance (target). For our first study, we simplified the problem by making sure that the structures of reference and target are similar (apart from interludes etc.). However, opera directors frequently decide to skip certain parts, shorten long finale sections, etc. Also, there may be different variants of an opera, with additional or alternative arias etc. (This is actually the case for Don Giovanni, of which there is a \textit{Prague} and a \textit{Vienna} version.) There are solutions for handling repetitions or omissions in offline alignment (e.g., \textit{JumpDTW} \cite{fremerey2010handling}); these could possibly be adapted for real-time application.

\section{Conclusion}
\label{sec:conclusion}

To summarize briefly, we have shown, using Mozart's \textit{Don Giovanni} as a showcase, that tracking live opera with state-of-the-art On-Line Time Warping (OLTW) is inaccurate, not robust, and leads to severe errors during transitions between scenes. We identified three specific categories or sources of error, proposed a partial solution by integrating various top-down control strategies based on audio event detectors, and showed experimentally that these help to strongly mitigate these problems.

We also saw that a number of problems remain, including improvised interludes, recitatives, and structural differences between performances. Thus, our immediate next research steps in this project will focus on audio-to-lyrics transcription and alignment, and on robust methods for quickly reacting to differences in section structure. However, we have to remind ourselves that opera, as an art form that has developed over several centuries, is considerably more complex and varied than the specific classical style represented by \textit{Don Giovanni}; extensive experiments with other operas from other periods will be needed to establish the robustness of any methods we will come up with.

\begin{acknowledgments}
This project has received funding from the European Union’s Horizon 2020 research and innovation programme under the Marie Sklodowsa-Curie grant agreement No. 765068 (Project MIP-Frontiers). Special thanks to Andreas Arzt for many fruitful discussions, and to Christopher Widauer for providing the Don Giovanni recordings from the Vienna State Opera.
\end{acknowledgments}

\bibliography{smc2020bib}

\end{document}